\documentclass[%
 reprint,
 amsmath,
 amssymb,
]{revtex4-1}

\usepackage{graphicx}
\usepackage{dcolumn}
\usepackage{bm}
\usepackage{epsfig,color,xspace,multirow,xr,bbold}
\usepackage[all]{xy}
\usepackage{setspace}
\usepackage{url}

\newcommand{\Aq}{{\rm Ala}_{4}}

\newcommand{\Rof}{R_{\operatorname{1--4}}}

\usepackage{hyperref} 

\hypersetup{
   colorlinks,
   menucolor=blue,
   linkcolor=black,
   citecolor=blue,
   urlcolor=blue
}

\begin{document}

\title{Accurate structure-based coarse-graining leads to consistent
  barrier-crossing dynamics}

\author{Tristan Bereau}
\email{bereau@mpip-mainz.mpg.de}
\author{Joseph F.~Rudzinski}

\affiliation{Max Planck Institute for Polymer Research, 55128 Mainz,
  Germany}

\date{\today}

\begin{abstract}
  Structure-based coarse-graining of molecular systems offers a
  systematic route to reproduce the many-body potential of mean
  force. Unfortunately, common strategies are inherently limited by
  the molecular-mechanics force field employed.  Here we extend the
  concept of multisurface dynamics, initially developed to describe
  electronic transitions in chemical reactions, to accurately sample
  the conformational ensemble of a classical system in equilibrium.
  In analogy to describing different electronic configurations, a
  surface-hopping scheme couples distinct conformational basins beyond
  the additivity of the Hamiltonian.  The incorporation of more
  surfaces leads systematically toward improved cross-correlations.
  The resulting models naturally achieve consistent long-time dynamics
  for systems governed by barrier-crossing events.
\end{abstract} 

\maketitle


The complex separation of length- and time-scales in soft-matter
systems calls for modeling strategies at different resolutions: from
quantum, to classical atomistic, to mesoscopic, to the continuum
scale~\cite{murtola2009multiscale, peter2009multiscale,
  peter2010multiscale, kamerlin2011coarse}.  Among them,
particle-based coarse-grained (CG) models, which try to remain close
to the chemistry while averaging over the faster degrees of freedom,
have offered significant insight into complex (bio)molecular
systems~\cite{nielsen2004coarse, voth2008coarse, brini2013systematic,
  noid2013perspective}.  These models shine through both an effective
reductionist approach to testing what interactions lead to reproducing
certain phenomena, and a significant computational speedup to tackle
systems prohibitively large at the atomistic scale.

Instead of targeting a potential energy surface, averaging over
degrees of freedom leads to a many-body potential of mean force
(MB-PMF)~\cite{noid2008multiscale, rudzinski2011coarse}.  While
several systematic methods exist to target the MB-PMF
~\cite{tschop1998simulation, noid2008multiscale, noid2008multiscale2,
  shell2008relative} their accuracy tends to be limited not by the
performance of the method, but rather by the molecular-mechanics terms
used to approximate the MB-PMF.  Several recent attempts have been
made at using more complex interaction terms~\cite{molinero2008water,
  sanyal2018transferable, john2017many}, illustrating the need for
more accurate models.  In this Letter, we present a unique strategy
for generating complex cross-correlations between interaction terms of
a force field, thereby accurately recovering the MB-PMF.  We analyze
the structural and dynamical properties of CG models in the limit of
accurately matching these cross-correlations.

Limitations of the molecular-mechanics force field have long been
addressed for chemical reactions by methods such as empirical valence
bond~\cite{warshel1980empirical}, its multisurface
extension~\cite{schmitt1998multistate}, and surface-hopping
schemes~\cite{tully1990molecular}.  In these approaches, reactions are
effectively decomposed into surfaces with distinct electronic
configurations, such as the two bonded states for a proton transfer.
The limitations of the force-field interaction terms are overcome by
coupling distinct potential energy surfaces (PESs) through
surface-hopping dynamics.



In the classical simulation community, researchers have coupled
distinct force fields to describe internal-state conversions of
coarse-grained units~\cite{Murtola:2009ys}, entanglement in polymer
melts~\cite{Chappa:2012}, and large-scale conformational transitions
of biomolecules~\cite{Knott:2014}.  In the case that the force-field
interaction functions can be expressed analytically as a function of a
continuous order parameter, e.g., in dissipative particle dynamics
models with local density-dependent
potentials~\cite{Pagonabarraga:2001}, no explicit hopping protocol is
required, as the dynamics along the continuous hypersurface of force
fields is well-defined by the normal integration scheme.  On the other
hand, if switching between force fields is discrete and, in
particular, if a timescale separation exists between force-field
transitions and the local motion of particles, Monte Carlo provides a
robust route for instantaneous switching between force fields.  Voth
and coworkers have recently laid out an elegant ``ultra
coarse-graining'' framework in this context, where conversions between
discrete internal states are modeled by stochastic transitions between
distinct force fields~\cite{dama2013theory, davtyan2014theory}.

In the present work, we expand upon previous efforts by considering
the common situation where significant coupling between local degrees
of freedom in the simulation model are essential for accurate modeling
of the structural ensemble.  To address this challenge, we draw an
analogy between electronic transitions and transitions between
conformational basins, in the context of surface-hopping techniques.
Instead of matching the PES due to different types of electronic
configurations, we aim to reproduce features of distinct
conformational basins of the underlying free-energy surface.  Thus, we
assign distinct force fields to conformations belonging to a given
basin, and hop between conformationally-dependent surfaces.  Rather
than hop between surfaces in a stochastic manner, we ascribe a
continuous-switching scheme.  In contrast to previous studies
employing discrete transitions between distinct force fields, the
transitions between local conformational basins considered in this
work occur on comparable timescales to the local dynamics,
discouraging the use of the Monte Carlo approaches.
Surface-hopping schemes for chemical reactions typically weight each
surface according to solutions of the Schr\"odinger equation,
effectively leading to a strong dependence on the relative energies.
Because we aim at reproducing the free energy of each conformational
basin, we require an integrated measure.  As such we define the
weights using a structural criterion: a metric depending collectively
on the instantaneous values of the order parameters governing each
interaction.
Despite retaining the standard molecular-mechanics form of individual
force fields, the conformationally-dependent surface hopping generates
complex cross-correlations between local degrees of freedom, as we
will demonstrate below.



\begin{figure}[htbp]
  \begin{center}
    \includegraphics[width=0.7\linewidth]{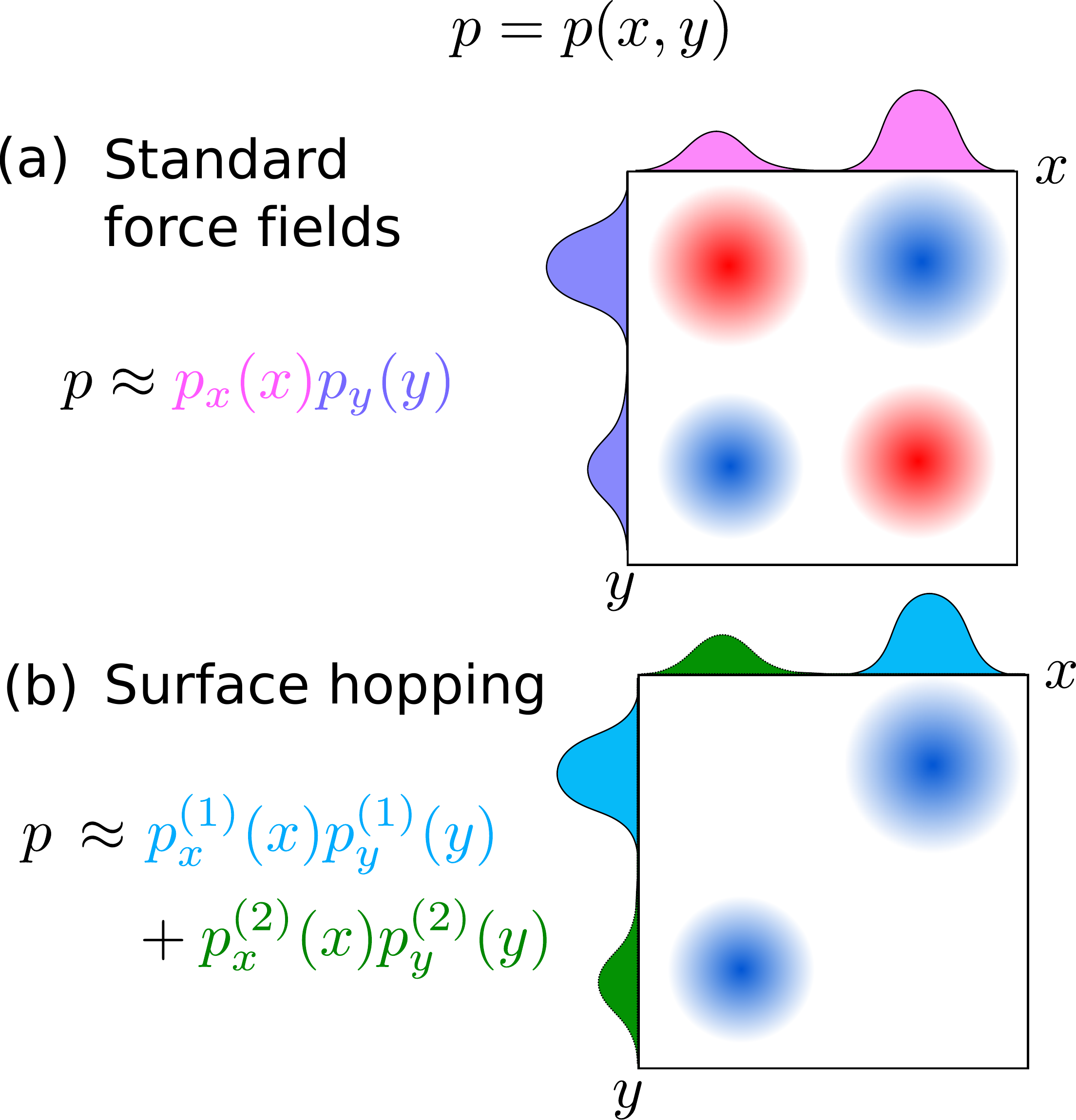}
    \caption{Consider a 2D potential $U=U(x,y)$ leading to the
      distribution $p=p(x,y)$ populated by two peaks (in blue). (a)
      Standard force fields apply a global separation of variables on
      the potential $U(x,y) \approx U(x) + U(y)$ such that $p \approx
      p_x(x)p_y(y)$ leading to two spurious peaks (in red). (b)
      Surface hopping retains the separation of variables, but
      determines a local force field per conformational basin.  }
    \label{fig:sketch}
  \end{center}
\end{figure}

\emph{Methodology.}  Instead of relying on a single force field, we
model a molecular system by means of $n$ force fields, each focusing
on specific conformational basins (Fig.~\ref{fig:sketch}).  The
splitting of conformational space is provided by a density-based
cluster analysis applied along the CG interaction terms of
interest~\cite{sittel2016robust}.  The reference trajectories are then
split according to cluster identities, allowing us to build $n$ force
fields using standard structure-based CG procedures (force-matching in
the present work).

The $i$-th cluster describes a subset of conformational space
projected down onto two CG interaction terms denoted $x$ and $y$
(Fig.~\ref{fig:sketch})---for instance a bond and bending angle, as in
the hexane application below.  We further define the cluster center,
${\bf \mu}^{(i)} = (\mu_x^{(i)}, \mu_y^{(i)})$, corresponding to a
local maximum of probability density across each variable.  Similarly,
we define the spatial extent of the cluster by means of its standard
deviation, ${\bf \sigma}^{(i)} = (\sigma_x^{(i)}, \sigma_y^{(i)})$.
We apply a linear transformation on the clusters to enhance their
isotropy: $\overline{\bf \sigma}^{(i)} = (\overline\sigma_x^{(i)},
\overline\sigma_y^{(i)})$.

Force field $i$ is characterized by its potential energy,
$U_i({\bf R})$, and corresponding force ${\bf f}_{i} ({\bf R}) = -
{\bf \nabla} U_i({\bf R})$.  Force field $i$ is assigned a coefficient
$w_i$ that weights its instantaneous contribution, such that the net
force on any particle is a weighted sum over all force fields
\begin{equation}
  \label{eq:wff}
  {\bf f} ({\bf R}) = \sum_{i=1}^n w_i {\bf f}_{i} ({\bf R}).
\end{equation}

The weight $w_i$ of force field $i$ is determined by the Euclidean
distance between the system's instantaneous configuration along the CG
interaction variables $(x,y)$ and the center of the $i$-th cluster $
d_i^2 = (x-\mu_x^{(i)})^2/\overline\sigma^{(i)}_x +
(y-\mu_y^{(i)})^2/\overline\sigma^{(i)}_y$, where the scaling by
${\overline\sigma^{(i)}_{x,y}}$ normalizes the contribution of each
interaction.  This distance is compared to the spatial extent of the
cluster through the norm $|\overline{\bf \sigma}^{(i)}|$.  If the
system is within the cluster's spatial extent its force field gets
full weight, otherwise we suppress the weight exponentially
\begin{equation}
  \label{eq:wi}
  w_i =
  \begin{cases}
    1 , \quad & d_i < |\overline{\bf \sigma}^{(i)}| \\
    \exp \left(- \frac{d_i - |\overline{\bf
          \sigma}^{(i)}|}\alpha \right) , \quad
    &\textup{otherwise}.
  \end{cases}
\end{equation}
The scaling factor $\alpha$ smoothly dampens the contribution of force
field $i$ to avoid numerical instabilities upon integrating the
equations of motion.  A value of $\alpha$ that is small compared to
the spatial extent of any cluster center avoids the blurring of the
different PESs.

The mixing of several force fields, as described in Eqn.~\ref{eq:wff},
can easily lead to unphysical behavior, even from a weak contribution
of a surface containing large restoring forces.  To avoid such a
behavior, we restrict the mixing as much as possible.  This is
achieved by defining the first $n-1$ surfaces that are localized to a
cluster center, while the last force field $n$ embodies the default
option.  This fallback surface is thus \emph{not} associated to any
cluster center, but instead parametrized from the rest of the
trajectory that has not yet been considered.  We compute the weights
$w_i$ (Eqn.~\ref{eq:wi}) for the first $n-1$ surfaces, keep only the
one with the largest contribution
$
w_l = {\max}_{i<n} w_i,
$
and assign the rest of the weight to the fallback surface,
$w_n = 1-w_l$.  As such, the surface mixing described in
Eqn.~\ref{eq:wff} is always limited to the closest cluster center and
the fallback force field.  When the system is far from any cluster
center, it relies solely on the fallback surface ($w_n = 1$).  Our
scheme directly hops between surfaces without rescaling velocities,
and thereby violates total-energy conservation.  In the canonical
ensemble the thermostat is capable of absorbing a certain amount of
energy violation~\cite{davtyan2014theory}, which we enhance by working
at high friction.

While the algorithm described above yields surface hopping, it doesn't
ensure the correct probabilities of sampling each surface.  To this
end, the ultra coarse-graining framework matches the transition rates
through a self-consistent optimization~\cite{davtyan2014theory}.  Here
we simply enforce the system to sample each surface $i$ such that the
time average $\langle w_i \rangle$, taken as a proxy of its canonical
probability, matches the target probability $p_i$---available upon
partitioning of the conformational space.  The matching is enforced by
locking the system that is currently visiting surface $i$ on that
force field until $\langle w_i \rangle \approx p_i$.  In practice, we
let the system escape from surface $i$ once $\langle w_i \rangle \geq
0.98~p_i$.  While we only constrain a lower bound on sampling each
surface, our experience so far indicates that it is sufficient to
recover the correct probabilities.

All simulation details are described in the Supplemental Material
(SM)~\footnote{See Supplemental Material {\bf [URL]} for simulation
  details; generation of the CG potentials; free-energy surfaces; and
  Markov state model analysis, which includes Refs.~\cite{Jain:2012,
    Jorgensen:1996, Berendsen:1987, Hess:2008, Izvekov:2005d,
    Izvekov:2005e, Noid:2007a, Noid:2008a, Mullinax:2009b,
    Mullinax:2010, Ellis:2011a, Prinz:2011b, humphrey1996vmd}}.  An
implementation of the surface-hopping scheme is available in {\sc
  ESPResSo++}~\cite{halverson2013espresso++}, as well as all
simulation and analysis scripts~\cite{gl_cg}.



\begin{figure}[htbp]
  \begin{center}
    \includegraphics[width=0.75\linewidth]{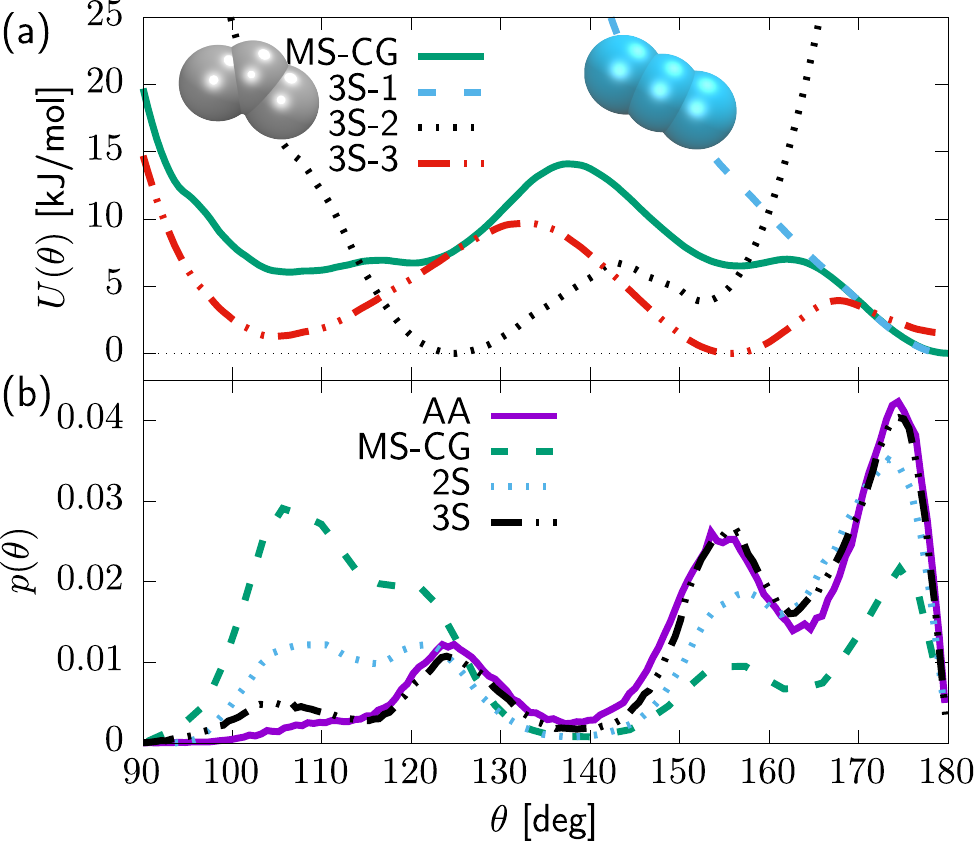}
    \caption{Bending-angle properties of the CG hexane molecule.  (a)
      Potential energy from force matching (MS-CG) and the three
      contributions of the 3-surface (3S) model.  Cartoons display
      representative structures of the low-energy states for the
      relevant surface.  (b) Probability distribution from the
      all-atom distribution projected onto CG variables (AA), force
      matching (MS-CG), and surface hopping with 2 (2S) and 3 (3S)
      force fields.}
    \label{fig:hexa}
  \end{center}
\end{figure}

\begin{figure}[htbp]
  \begin{center}
    \includegraphics[width=0.9\linewidth]{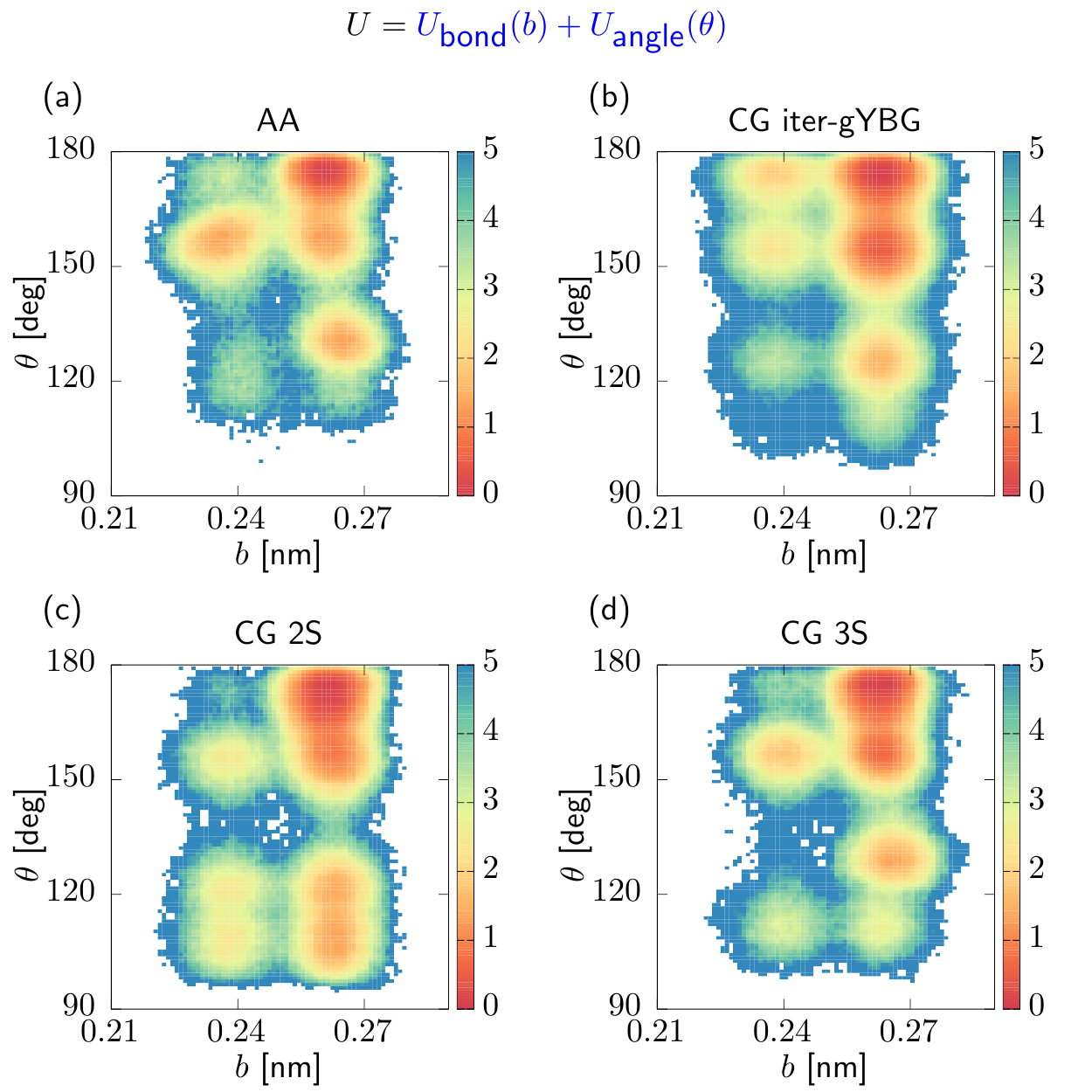}
    \caption{Free-energy surfaces of the hexane molecule as a function
      of the bond, $b$, and bending angle, $\theta$.  (a) Reference
      surface; (b) CG iterative gYBG, (c) CG 2-state and (d) CG
      3-state surfaces.  Surface hopping couples bond with bending
      angle.  Free energies expressed in $k_{\rm B}T$.}
    \label{fig:hexfes}
  \end{center}
\end{figure}

\emph{Hexane.}  We first consider a single hexane molecule in vacuum
coarse-grained to 3 beads---a challenging case despite its apparent
simplicity.  The CG potential employed bonded interactions between
subsequent pairs of beads along the chain and an angle-bending
interaction between the three beads.  This CG model was first
described in R\"uhle \emph{et al.}~\cite{ruehle2009versatile}.  The
force-matching-based multiscale coarse-graining (MS-CG) method applied
to the reference all-atom (AA) trajectory led to significant
structural discrepancies, as seen in the 1-dimensional bending-angle
distribution (Fig.~\ref{fig:hexa}b).  MS-CG overpopulates small-angle
states ($100^\circ < \theta < 120^\circ$), while underpopulating the
high-angle states ($\theta > 150^\circ$).  Rudzinski and Noid later
demonstrated that these discrepancies arise due to bond-angle
cross-correlations that cannot be reproduced with the
molecular-mechanics interaction set~\cite{Rudzinski:2014}.  They
applied an iterative generalized Yvon-Born-Green (iter-gYBG) scheme to
reproduce the independent bond and bending-angle reference AA
distributions, albeit at the cost of accuracy in the
cross-correlations.  The discrepancies in the cross-correlations
between bond and bending angle generated by these models is
illustrated by the free-energy surfaces (FESs) in
Fig.~\ref{fig:hexfes}.  The AA model displays a complex surface made
of four major minima, located asymmetrically on the surface.  The
symmetry of the iter-gYBG model, on the other hand, clearly
illustrates the additivity of the interactions in the Hamiltonian: all
large-bond states are more populated than the small-bond states,
irrespective of the angle.

A clustering of conformational space in two surfaces (i.e., 2S model)
leads to: (2S-1) the highest angle state coupling to the large bond
state (cluster center: $b = 0.26$~nm, $\theta = 170^\circ$) and the
fallback surface (2S-2).  Fig.~\ref{fig:hexa}a shows how a 3-state CG
model discriminates between: (3S-1) the highest angle
state---identical to 2S-1; (3S-2) two intermediate angles
($\theta \approx 125^\circ$, $\theta \approx 155^\circ$) with large
bond; and the fallback surface (3S-3): an intermediate and a low angle
state ($\theta \approx 155^\circ$, $\theta \approx 105^\circ$) around
both small and large bond states.

Fig.~\ref{fig:hexa}b displays the bending-angle canonical
distributions for the different CG models.  We find that 2S and 3S
systematically refine the agreement with the AA distribution as
compared to MS-CG: they lower the artificially large populations of
small-angle states and increase the artificially-low populations of
high-angle states.  The 3S model reproduces the AA angle distribution
remarkably accurately, with only a slightly-low population around
$\theta \approx 105^\circ$.  In addition, the interaction potentials
become more localized: while the MS-CG potential is extremely broad, the
different 3S potentials are better confined.  

The effect of the surface hopping technique is even more apparent in
the FES.  The 2S model demonstrates significant improvement relative to the 
standard molecular-mechanics force field by more accurately
representing the heterogeneous populations of high-angle states.  The
3S model further corrects the populations, especially for the
low-angle states.  

A Markov state model analysis~\cite{Noe:2008, Bowman:2014} of both the
AA and different CG models yielded a lag time too close to the longest
timescales to make use of the results, indicating that the dynamics
are governed by diffusive behavior and lack timescale separation
between the conformational basins.


\emph{Tetraalanine}: As a second system, we consider $\Aq$, a
tetraalanine peptide made of 52 atoms solvated in water,
coarse-grained to only four beads ~\cite{Rudzinski:2014b}.  Each bead
was placed at the position of the alpha carbon on the peptide
backbone.  The CG force field employed bonded interactions between
subsequent pairs of sites along the peptide chain, two angle-bending
interactions, $\theta$, a dihedral interaction, $\psi$, and an
additional effective bond between the terminal beads of the chain,
$\Rof$.

\begin{figure}[htbp]
  \begin{center}
    \includegraphics[width=0.9\linewidth]{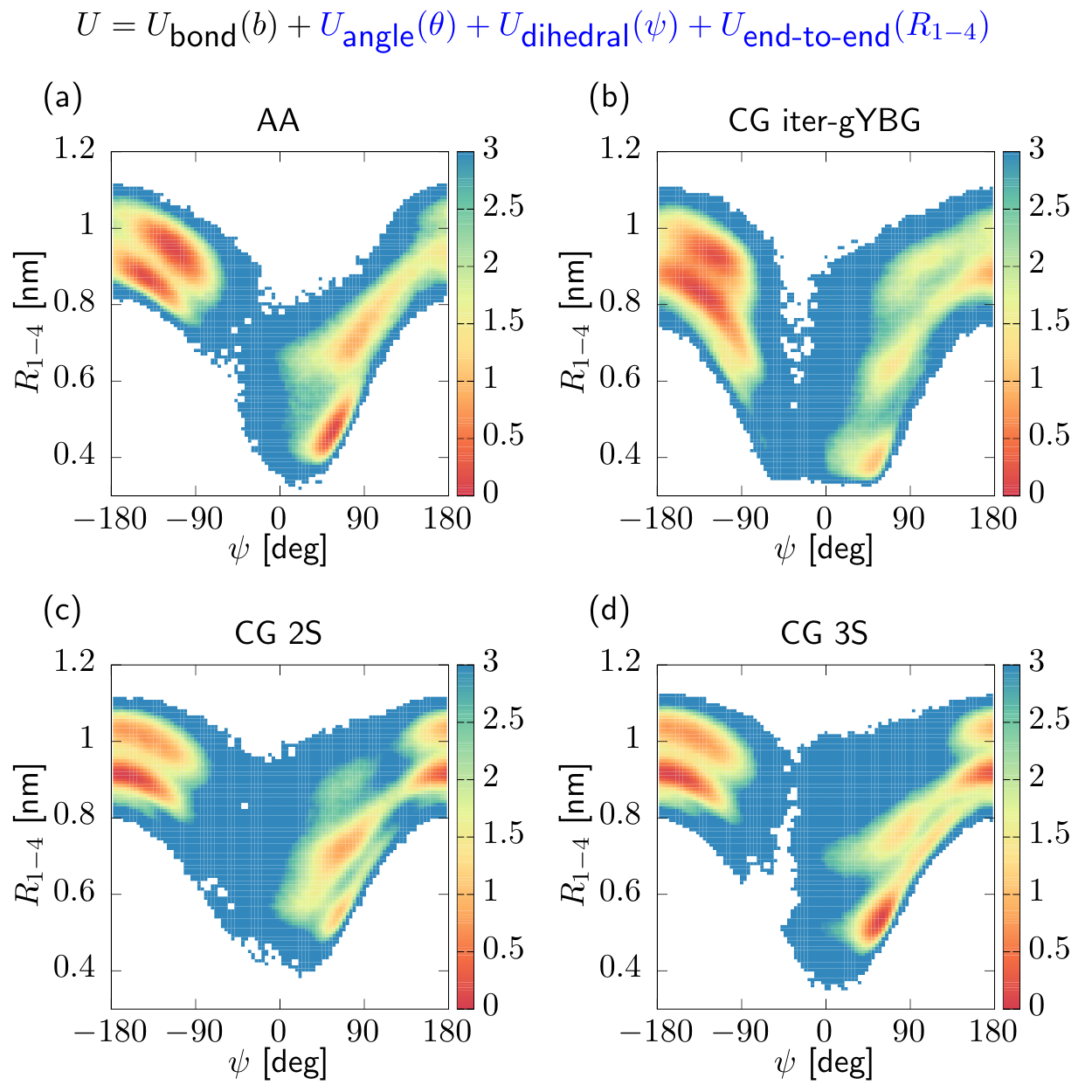}
    \caption{Free-energy surfaces of $\Aq$ as a function of the
      dihedral, $\psi$, and end-to-end distance, $\Rof$.  (a)
      Reference surface; (b) CG iterative gYBG, (c) CG 2-state and (d)
      CG 3-state surfaces.  Surface hopping couples the bending angle,
      dihedral, and the effective $\Rof$ interaction.  Free energies
      expressed in $k_{\rm B}T$.}
    \label{fig:alafes}
  \end{center}
\end{figure}

Both MS-CG and iter-gYBG models displayed two notable discrepancies on
the FES (Fig.~\ref{fig:alafes} and SM): a spurious region of
intermediates ($\psi \approx 90^\circ$, $\Rof \approx 0.9$~nm) and the
extended state stabilized at too large dihedrals ($\psi \approx
-90^\circ$, $\Rof \approx 0.6$~nm).  Surface-hopping models that
couple $\theta$, $\psi$, and $\Rof$ using 2 and 3 surfaces suppress
both regions.  Further, a 4S model (SM) shows structural accuracy on
par with 3S.  The results highlight the capability of the
surface-hopping scheme to introduce cross-correlations beyond the
additivity assumption of the Hamiltonian.  This improvement is due not
only to the introduction of cross-correlations between interactions,
but also to the simplification of the target surface when determining
each force field.  In particular, we have found that as the number of
surfaces increases, the distributions within each basin become
increasingly unimodal, resulting in very simple interaction potentials
and systematically improving the accuracy of the model.

\begin{figure}[htbp]
  \begin{center}
    \includegraphics[width=0.9\linewidth]{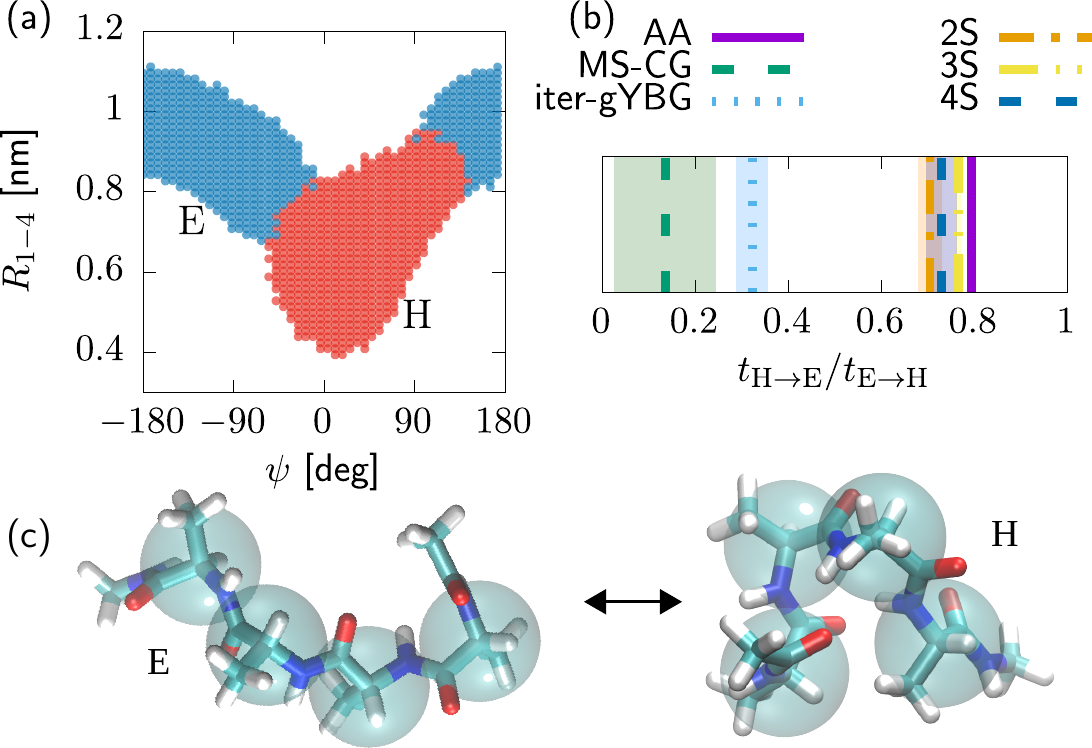}
    \caption{(a) Basin decomposition between helical (H) and extended
      (E) states.  (b) Ratio of mean-first-passage times between the
      helical and extended states,
      $t_{\mathrm{H} \rightarrow \mathrm{E}} / t_{\mathrm{E}
        \rightarrow \mathrm{H}}$.  (c) Cartoon representations of the
      extended and helical states.}
    \label{fig:alamfpt}
  \end{center}
\end{figure}

Beyond structural properties, the dynamics also show significant
improvements.  We monitor the transition kinetics between the helical
(H) and extended (E) metastable basins (Fig.~\ref{fig:alamfpt}a and
c), identified from a Markov state model analysis of the reference AA
simulation~\cite{Rudzinski:2014b, rudzinski2016communication}.  We
focus on \emph{ratios} of mean-first-passage times to factor out any
homogeneous speedup factor due to coarse-graining.  Compared with the
MS-CG and iter-gYBG models, the ratios of mean-first-passage times
converge more consistently over a wider range of characteristic
timescales, indicating better-defined kinetic boundaries between
conformational basins.  The surface hopping schemes also yield
significantly better agreement with the AA result
(Fig.~\ref{fig:alamfpt}b).  We observe a systematic improvement for
the different surface-hopping models, where both 3S and 4S lie almost
within the error bars of the reference AA observable.  The results
indicate that not only are the free-energy barriers well reproduced,
the diffusive behavior in the different basins is consistently sped
up.

Achieving consistent long-time dynamics of a tetraalanine peptide
solvated in water using only four beads is remarkable, given the
tendency of CG models to display severe kinetic
discrepancies~\cite{peter2009multiscale}.  In previous studies,
kinetically-consistent CG models have only been systematically
achieved using a Mori-Zwanzig formalism, where a generalized Langevin
equation introduces a computationally-expensive memory kernel to
account for the degrees of freedom coarse-grained
away~\cite{hijon2010mori}.  Here, our results demonstrate that a
simple model solely parametrized against structural properties can
quantitatively reproduce the long time-scale dynamics.  Indeed, the
model's accurate description of the free-energy barriers enforces the
barrier-crossing dynamics, akin to Marcus theory for electron-transfer
reactions~\cite{marcus1956theory}.  On the other hand, local diffusion
within a conformational basin remain inconsistent because of reduced
molecular friction~\cite{rudzinski2016communication}.  Our findings
generalize beyond coarse-graining: surface-hopping models offer a
systematic method to achieve accurate barrier-crossing dynamics, but
the quality of the local diffusion is inherently limited by the
physics of the model.  Empirical valence bond models showed a strong
dependence of the quality of the individual surfaces on excess proton
dynamics~\cite{wu2008improved}.  Systems whose kinetics are dominated
by activated processes can thus be accurately characterized with
remarkably simple force fields, unlike in the diffusive regime.

\section*{Acknowledgments}

We thank Denis Andrienko, Burkhard D\"unweg, Kurt Kremer, and Clemens
Rauer for insightful discussions.  This work was supported in part by
the TRR 146 Collaborative Research Center of the Deutsche
Forschungsgemeinschaft (DFG), the Emmy Noether program to TB and a
postdoctoral fellowship from the Alexander von Humboldt foundation to
JFR.

\bibliography{biblio} 

\end{document}